\documentclass[a4paper]{article}

\usepackage{icrc2013}

\title{The problem of small angular scale structure in the cosmic ray anisotropy data}

\shorttitle{Small angular scale anisotropies}

\authors{L. O'C. Drury}

\afiliations{
Dublin Institute for Advanced Studies\\ 
School of Cosmic Physics\\ 
31 Fitzwilliam Place\\ 
Dublin 2\\
Ireland}

\email{ld@cp.dias.ie}

\abstract{
Recent observations have revealed structure on small angular scales in the anisotropy data of cosmic rays at multi TeV energies.  Even though the absolute amplitudes are very small, these effects are somewhat surprising and a wide range of possible causes have been discussed.    A possible origin associated with heliospheric electric fields is proposed.}


\begin{document}

\maketitle

\section{Introduction}
Our knowledge of directional variations (usually referred to as anisotropies) in the flux of cosmic rays reaching the Earth at TeV energies and above has greatly improved in recent years, largely because of the need to control the charged cosmic ray background in experiments searching for gamma rays or neutrinos.   Among others we now have excellent data from Super-Kamiokande, Tibet III, Milagro, ARGO-YGB, IceCube and EAS-Top, \cite{An1,An2,An3,An4,An5,An6, An7,An8}.
A surprising feature of these studies is that there is clear evidence for small-angular scale structure in the anisotropy data, including localised `hot-spots' as first reported by Milagro \cite{An3}.  Small here means on angular scales of a few to twenty degrees.   While a number of authors have recently discussed interpretations of the large angular scale anisotropies in terms of cosmic ray source and transport models (e.g. \cite{EW, Biermann} and references therein) little attention, with the notable exceptions of \cite{Silk,LD,DL,GS}, appears to have been paid to the small-scale structure which, as pointed out in \cite{DA}, is quite puzzling and hard to understand.  As noted in \cite{An3} the more prominent of the Milagro hot-spots appears to be directionally coincident with the heliotail, the extended region of shocked solar wind and interstellar medium trailing behind the solar system as it moves through the local interstellar medium at $25\,\rm km\,s^{-1}$ \cite{Bzowski}.

\section{Current explanations}

Simple diffusion models of CR propagation imply a dipole anisotropy, no matter how complex the distribution of sources and sinks or how time-variable the system, and thus are clearly inadequate to explain the observations.  More elaborate pitch-angle scattering models can do a bit better, but generally imply gyrotropic distributions which are symmetric about the mean magnetic field direction and do not naturally produce structure on the small scales observed.  Focussed transport and leakage through magnetic mirrors can produce field-aligned beams of small angular size, but only one `spot' along the mean field \cite{DA}.  More complex plasma process have been considered in \cite{Malkov}. 

Albedo neutron models, where the anisotropy signal (which appears to be hadronic in nature) is attributed to  secondary neutrons produced in localised near-by gas targets, have no difficulty with the small scale structure, but fail because they require unreasonably high target densities in close proximity to the sun which are excluded by astronomical observations
\cite{DA}.

A recent fascinating but highly speculative suggestion is that the anisotropy might be due to neutral quark matter lumps, so-called strangelets \cite{Silk}, produced through interactions in molecular clouds.

The apparent connection to the heliotail has prompted suggestions of a link to acceleration by magnetic reconnection \cite{LD} or anomalous scattering  \cite{DL} in the heliosheath region although it would be surprising for such processes to work on the multi TeV energy scale.

In a very interesting recent paper Giacinti and Sigl \cite{GS} argue that the small scales can naturally arise from scattering on local magnetic field structures combined with the presence of a large-scale anisotropy and that there is in fact no problem.  While this is physically correct and a possible explanation it need not be the only one.

\section{An alternative explanation}

It is usually assumed that the weak but complex anisotropy structure observed at TeV energy scales must be produced by processes operating on that energy scale.  However because the amplitude of the anisotropy is very small, of order $10^{-4}$, the same signal could equally well be generated by processes operating at energies four orders of magnitude smaller, but affecting all the particles coming from a particular direction   Specifically if all particles arriving from a given direction have their energy shifted slightly by passing through a retarding or accelerating electric field it is possible to produce a complex anisotropy signal at the $10^{-4}$ level with processes operating on length and energy scales that are four orders of magnitude smaller.  Potentials as low as $100\,\rm MV$ could in this way suffice to generate a signal at the $10^{-4}$ level on TeV energy scales.    This is the key idea explored in this note.  

In most discussions of cosmic ray transport we ignore electric fields and assume that only magnetic fields are important, the justification being the very high conductivity of space plasmas that shorts out any large-scale electric field.  But this is of course frame dependent.  Even if the electromagnetic field is purely magnetic in the plasma rest frame, a velocity boost into a moving frame will induce an electric component.  Specifically, in most regions of space the ideal MHD condition implies that 
\begin{equation}
E + V \times B = 0
\end{equation}
where $E$ is the electric field, $B$ the magnetic field and $V$ the plasma velocity.  Thus a moving plasma always has an associated induction field of $-V\times B$. For magnetised particles, that is on scales larger than the particle gyroradius  this is a largely formal result.   The energy gained by a charged particle from the electric field during one half of a gyration is exactly cancelled by that lost in the second half, and the net effect gives rise to the familiar $E\times B$ drift of charged particles with the bulk plasma (which is normally simply thought of as the particles being stuck to the field lines and the field lines being advocated with the plasma). It is more interesting in the case of unmagnetised particles where the field can be seen as a real accelerating or retarding field.

Specifically, in the case of the heliosphere the characteristic length scale is of order a hundred AU or $1.5\times 10^{13}\,\rm m$.  This is about the same as the gyro-radius of a $1\,\rm TeV$ proton  in typical interstellar (and outer heliosphere) magnetic fields of a few nT.  Thus above about a TeV cosmic ray particles are expected to penetrate the heliosphere with relatively little deflection.  Such particles arriving at the Earth will thus have seen an effective potential shift due to induction fields in the heliosphere of 
\begin{equation}
\int -V \times B \cdot ds
\label{eq1}
\end{equation}
where the integral is taken along the trajectory of the incoming particle out from Earth through the heliosphere and into the local ISM.  It is important that all these particles arriving in a given direction and energy band will have essentially the same trajectory through the heliosphere.  If the trajectories diverged rapidly, as they do at lower energies, the effect would be wiped out.  To order of magnitude the potential shift, assuming a heliospheric length scale of a hundred AU, a typical velocity scale of between $10^4\,\rm m\,s^{-1}$ and $10^{5}\,\rm m\,s^{-1}$, and a $1\,\rm nT$ magnetic field, will be $100\,\rm MV$ to $1\,\rm GV$, similar to that characteristic of solar modulation (and of course for the same reason).  Thus to order of magnitude we should indeed expect signatures of heliospheric electric field structure in the TeV anisotropy data at the level of $10^{-4}$ as observed.   If these fields are located in the outer heliotail the associated time-scales will be of order 100 years or more, sufficiently long for the anisotropy spots to appear static in currently running experiments as appears to be the case.

This model has a number of attractive features.  It relies only on very simple basic physics; it naturally explains the low amplitude of the signal and the fact that it is observed at energies above a TeV; and it explains the correlation with heliospheric structure and is capable (depending on the fields in the heliotail) of generating multiple hotspots and complex non-gyrotropic structures.
The next step is clearly to take a detailed MHD model of the heliosphere, evaluate the integral (\ref{eq1}) along self-consistently calculated particle trajectories and compare the results with the actual data.   One interesting prediction of this model is that the signal in the electrons (if it could be observed) should be exactly in anti-phase with that in the protons; a hotspot in the proton data should be a cold spot in the electrons and vice-versa.  

\section{Conclusions}

It is suggested that the small angular scale TeV anisotropy data observed in the cosmic rays may in part be a reflection of the electric field structure of the outer heliosphere.  If this is the case, it offers an interesting way to test and constrain models of the interaction between the solar wind and the local interstellar medium.  The mechanism proposed here is completely distinct from, and conceptually much simpler than, those suggested by Lazarian and  Desiati \cite{LD,DL} although both seek to relate the small-scale structure observed in the anisotropy data to heliospheric structure. 

It should be emphasised that there must be other contributions to the anisotropy, in particular the large scale components are surely a reflection of source distributions and global transport in the Galaxy as well as the Compton-Getting effect due to the motion of the solar system.  The point of this note is simply to point out a simple and plausible mechanism whereby there could be a heliospheric component to the small-scale signal even at TeV energies.

\section{Acknowledgment}

I thank Andrew Taylor for helpful comments on an earlier draft of this note and for drawing my attention to the important work by Giacinti and Sigl \cite{GS}.  Correspondence with G. Giacinti  and M. Santander is also gratefully acknowledged.


\begin{thebibliography}{}

\bibitem{An1} G. Guillian et al, PRD 75 (2007) 062003.
\bibitem{An2} M. Amenomori et al, Science 314 (2006) 439.
\bibitem{An3} A. A. Abdo et al, PRL 101 (2008) 221101.
\bibitem{An4} A. A. Abdo et al, ApJ 698 (2009) 2121.
\bibitem{An5} S. Vernetto et al, ArXiV:0907.4615
\bibitem{An6} R. Abassi et al, PRL 104 (2010) 161101.
\bibitem{An7} R. Abassi et al, ApJ 740 (2011) 16.
\bibitem{An8} M. Aglietta et al, ApJ 692 (2009) L130.

\bibitem{Bzowski} M. Bzowski et al, ApJS 198 (2012) 12.


\bibitem{DA} L. O'C. Drury and F. A. Aharonian, Astroparticle Physics 29 (2008) 420.
\bibitem{Malkov} M. A. Malkov et al, ApJ 721 (2010) 750.
\bibitem{Silk} K. Kotera, M. Perez-Garcia and J. Silk, arXiv:1303.1186
\bibitem{LD} A. Lazarian and P. Desiati, ApJ 722 (2010) 188.
\bibitem{DL} P. Desiati and A. Lazarian arXiv:1111.3075 ApJ 762 (2013) 44.

\bibitem{GS} G. Giacinti and G. Sigl, PRL 109 (2012) 071101.

\bibitem{Biermann} P. Biermann et al, ApJ 768 (2013) 124.
\bibitem{EW} A. D. Erlykin and A. W. Wolfendale, arXiv:1303.2889

\end{thebibliography}
\end{document}